\documentclass[11pt]{article}
\usepackage{amssymb}
\usepackage{mathrsfs}
\usepackage{euscript}

\def \beq{\begin{equation}}
\def \eeq{\end{equation}}
\def \beqa{\begin{eqnarray}}
\def \eeqa{\end{eqnarray}}
\def \Dgadot{\bar D^{\dot\alpha}}
\def \Ddadot{\bar D_{\dot\alpha}}
\def \Vaa{V^{\alpha\dot{\alpha}}}
\def \Dga{D^\alpha}
\def \Dda{D_\alpha}

\begin{document}
\author{ \bf Jerzy Szwed \rm \footnote{\tt email: szwed@cpt.univ-mrs.fr}
\\
Centre de Physique Th\'eorique, CNRS Luminy Case 907, 
\\
13288 Marseille Cedex 09, France 
\\
and
\\
Institute of Physics, Jagellonian University,
\\ 
Reymonta 4, 30-059 Krak\'ow, Poland
}
\title{\small CENTRE DE PHYSIQUE TH\'EORIQUE \footnote{Unit\'e Mixte de Recherche du CNRS et des Universit\'es de Provence, de la M\'editerran\'ee et du Sud Toulon-Var - Laboratoire affili\'e a la FRUNAM - FR 2291}
\\
CNRS-Luminy, Case 907
\\
13288 Marseille Cedex 9
\\
FRANCE
\vskip1cm
\large \bf The "square root" of the Dirac equation and solutions on superspace. \footnote{Work supported by
the European Commission contract ICA1-CT-2002-70013/INCO Strategic action on training and excellence, 5FP}}

%\pacs{}
\maketitle
%\vskip 1cm 
%\noindent Key-Words: supersymmetry
%\vskip 5mm
%\noindent 
CPT-2004/P.008

%-----    Abstract    ------------------------------------------------
\begin{abstract}
The "square root" of the Dirac operator derived on the superspace is used to construct supersymmetric field equations. In addition to the recently found solution - a vector supermultiplet - it is  demonstrated how other supermultiplets follow as solutions: a set of chiral and antichiral superfields, containing two spin $0$ and two spin $1/2$ component fields and a another set of two spin $3/2$ and two real, spin $1$ component fields.  These  supermultiplets are shown to obey appropriate (massless) equations of motion. The "square root" of the Dirac equation yields thus a complete set of fields and their equations necessary to construct renormalizable supersymmetric theories. The problem of masses and interaction is also discussed.
\end{abstract}

\newpage

{\bf 1. Introduction.}

The idea to take the "square root" of the Dirac operator follows directly from the analogous procedure performed by Dirac on the Klein-Gordon operator \cite{Dirac}. Whereas in the first case the motivation was to linearize the operator in space-time derivatives, the form of the supersymmetry algebra \cite{supersymmetry} suggested that repeating this procedure would lead to the operator linear in supersymmetry generators or equivalently - spinorial derivatives. Such construction was presented some time ago \cite{JS86} together with a set of supersymmetric field equations which result when acting with the "square root" operator on superfields. 

To recall this construction in short let me write the Dirac equation in two-component notation and chiral representation:
\begin{equation}
\label{Dirac m}
-\left(
\begin{array}{cc}
i\bar\sigma^{\mu\,\dot\alpha\alpha}\partial_\mu & m \\
m & i\sigma^\mu{}_{\alpha\dot\alpha}\partial_\mu
\end{array}
\right)
\left(
\begin{array}{c}
\varphi_\alpha \\ \bar\chi^{\dot\alpha}
\end{array}
\right)
\; \equiv \;
{\mathcal D}\left(
\begin{array}{c}
\varphi_\alpha \\ \bar\chi^{\dot\alpha}
\end{array}
\right)
\; = \; 0.
\end{equation}
The Lorenz indices are denoted here  by $\mu, \nu, \lambda$ and $\rho$, the spinor indices by $\alpha$ and  $\beta$.

I am looking for the operator $\mathcal{S}$
 satysfying:
\begin{equation}
\label{prop1}
\mathcal{S^\dag\!S = D}.
\end{equation}
The solution proposed in Ref. \cite{JS86} is:
\begin{equation}
\label{S:solution}
\mathcal{S} \; = \;
{1\over\sqrt 2}
\left(
\begin{array}{rr}
\Dga & -\Ddadot \\
\Dgadot & \Dda
\end{array}
\right)
\end{equation}
where the spinorial derivatives are defined on the superspace as:
\begin{eqnarray}
\Dda & = & \;\;\partial/{\partial\theta^\alpha} +
i\sigma^\mu{}_{\alpha\dot\alpha}\bar\theta^{\dot\alpha}\partial_\mu,
\nonumber \\
\Ddadot & = & \!-{\partial}/{\partial\bar\theta^{\dot\alpha}}
-i\theta^{\alpha}\sigma^\mu{}_{\alpha\dot\alpha}\partial_\mu.
\end{eqnarray}
Indeed, using the anticomutation relations:
\begin{eqnarray}
\label{algebra}
\left\{\Dda,D_\beta\right\}
& = &
\left\{\Ddadot,\bar D_{\dot\beta}\right\}
= 0,
\nonumber \\
\left\{\Dda,\bar D_{\dot\beta}\right\}
& = & -2i\sigma^\mu{}_{\alpha\dot\beta}\partial_\mu
\end{eqnarray}
we get 

\begin{equation}
\label{Dirac M}
\mathcal{S^\dag \! S} =
-\left(
\begin{array}{cc}
i\bar\sigma^{\mu\dot\alpha\alpha}\partial_\mu & M \\
M & i\sigma^\mu{}_{\alpha\dot\alpha}\partial_\mu
\end{array}
\right)
\end{equation}
where a scalar, hermitian operator 
\begin{equation}
\label{M:definition}
M \; = \; -\frac14\left(D\!D +\bar D\!\bar D\right)
\end{equation}
appears instead of the mass $m$. The operator $\mathcal S$ is thus the 
solution to our problem  on the space of superfields $\Lambda$
which satisfy
\begin{equation}
\label{additional}
M\;\Lambda \; = \; m \; \Lambda.
\end{equation}

Note another solution to the problem: 
\beq
\label{Stilde}
{\mathcal{S'}} \; = \;
{1\over\sqrt 2}
\left(
\begin{array}{rr}
\Dgadot & \Dda \\
-\Dga & \Ddadot
\end{array}
\right) .
\eeq

Acting with the operator $\mathcal S$ (or $\mathcal S'$) on a superfield we are able to construct a free field equation - the "square root" of the Dirac equation. The simplest 2 choices of superfields are \cite{JS86}:
\beq
\label{F}
 F \; = \;
\left(
\begin{array}{r}
 W_\alpha \\
\bar{ \cal H}^{\dot\alpha}
\end{array}
\right).
\eeq
and
\beq
 B\;=\;\left(
\begin{array}
[c]{c}%
 \Phi\;\\
\Vaa%
\end{array}
\right)\label{B}%
\eeq
leading to the equations:
\beq
\label{seq}
 {\mathcal{S}} F = 0, \quad   {\mathcal {S}} B = 0,. 
\eeq
One can easily verify that the operator $\mathcal{S'}$ leads to the equivalent set of field equations. It is also obvious that due to Eq. (\ref{prop1}) both superfields $F$ and $B$ satisfy the Dirac equation 
\beq
{\mathcal{D}} F = 0, \quad
 {\mathcal{D}} B = 0.
\eeq
 Recently the equations:
\beq
{\mathcal{S}} F = 0
\eeq
(together with the condition $M F = m F$) were studied and solved in Ref. \cite{BH}. In the simplest case when $W_\alpha = \cal H_\alpha$,
the solution was found to be the Maxwell  supermultiplet: 
\beqa
\label{structure}
W_\alpha  =  & - & i\lambda_\alpha(y) + \left[\delta_\alpha{}^\beta d(y)
-\frac{i}{2}\left(\sigma^\mu\bar\sigma^\nu\right)_\alpha{}^\beta (\partial_\mu w_\nu (y) - \partial_\nu w_\mu(y))\right]\theta_\beta
\nonumber \\
& + & \theta\theta\sigma^\mu{}_{\alpha\dot\alpha}\partial_\mu\bar\lambda^{\dot\alpha}(y)
\eeqa
with $y^\mu = x^\mu +i\theta \sigma^\mu \bar \theta$.
The massless component fields $w_\mu(x)$ and $\lambda_\alpha(x)$   satisfy the  Maxwell and Dirac equations respectively and $d = const$.

\vskip1cm
{\bf 2. The equations and their solutions.}

In this paper I study the other set of equations (\ref{seq}):
\beq
{\mathcal{S}} B = 0,
\eeq
the superfield $B$ satisfying in addition the condition (\ref{additional}):
In terms of (in general complex) component superfields the equations read:
\beqa
\label{explicit}
D^\alpha \Phi - \bar D_{\dot\alpha} V^{\alpha\dot{\alpha}} & = & 0 ,
\nonumber \\
\bar D^{\dot\alpha} \Phi + D_\alpha V^{\alpha\dot{\alpha}} & = & 0
\end{eqnarray}
and
\beqa
M \Phi & = & m \Phi ,
\\
\nonumber
M \Vaa & = & m \Vaa .
\eeqa
Multiplying the first Eq. (\ref{explicit}) by $\Dda$ and making use of the second Eq. (\ref{explicit})
I obtain 
\beq
\label{constant}
M \Phi = m \Phi = 0.
\eeq
Out of two possibilities suppose first $\Phi = 0$ and the mass $m$  arbitrary. The Eqs (\ref{explicit}) simplify and it is easy to show in this case that
\beq
\label{DSQ}
D^2 \Vaa = \bar D^2 \Vaa = 0,
\eeq
which implies:
\beq
M \Vaa = m \Vaa = 0.
\eeq
If we are interested in non-zero superfields, the mass $m$ has to vanish, $m = 0$. The other possibility, $m\,=\,0$ and $\Phi$ arbitrary, leads to similar conclusion.
Indeed, a new constraint on $\Phi$ can be obtained by acting with the anticomutator 
$\left\{D_\alpha,\bar D_{\dot\beta}\right\}$
on the superfield $\Phi$ and using Eqs. (\ref{explicit}):
\beq
\left\{D_\alpha,\bar D_{\dot\beta}\right\} \Phi 
 =  -2i\sigma^\mu{}_{\alpha\dot\beta}\partial_\mu \Phi
 =  2M V_{\alpha \dot \beta} = 0.
\eeq
The above equality means that $\Phi$ is constant in space-time and depends only on $\theta$ and $\bar \theta$. Taking into account the condition (\ref{constant}), the most general form of $\Phi$ is:
\beq
\label{fi c}
\Phi_c = c_1 + c_2^{\alpha} \theta_{\alpha} + \bar c_{3\dot \alpha}\bar \theta^{\dot \alpha} + 
c_{4\mu}\theta \sigma^{\mu} \bar \theta+ c_5(\theta \theta - \bar \theta \bar \theta)
\eeq
with constant $c_1, c_2, ..., c_5$.
One further notices that the equations (\ref{explicit}) are invariant under the simultaneous shift:
\beqa
\label{shift}
\Phi  & \rightarrow & \Phi^{'}  =  \Phi + \Phi_c ,
\nonumber \\
\Vaa  & \rightarrow & V^{'\alpha \dot \alpha}  =  \Vaa + \Vaa_c
\eeqa
where
\beq
\Vaa_c = c_2^{\alpha}\bar \theta^{\dot \alpha} - \theta^{\alpha} \bar c^{\dot \alpha}_3 
+ c_{4\mu} \bar \sigma^{\mu \dot \alpha \alpha}(\theta \theta - \bar \theta \bar \theta)
+c_5 \theta^{\alpha}\bar \theta^{\dot \alpha}.
\eeq
I can perform this shift so that
\beq
\Phi = 0.
\eeq
Eqs. (\ref{explicit}) reduce then to:
\beq
\label{reduced}
\Ddadot \Vaa = \Dda \Vaa = 0.
\eeq
I first express the bi-spinor superfield $\Vaa$ through the vector superfield $V_{\mu}$:
\beq
\Vaa = \sigma^{\mu\dot\alpha \alpha} V_{\mu}
\eeq
Acting with the anticomutator $\left\{D_\alpha,\bar D_{\dot\beta}\right\}$
on the superfield $\Vaa$ one notices that 
the superfield $V_{\mu}$ is divergenceless:
\beq
\label{div}
\partial^{\mu} V_{\mu}=0.
\eeq
and due to Eq.(\ref{DSQ}):
\beq
\label{DDV}
D^2 V_{\mu} = \bar D^2 V_{\mu} = 0.
\eeq
To find the general solution to Eqs. (\ref{reduced}, \ref{div}, \ref{DDV}) let me 
expand $V_{\mu}$ in terms of component fields:
\beqa
V_{\mu} & = & a_{\mu}(x) + \sqrt{2} \theta \psi_{\mu}(x)  +
 \sqrt{2} \bar \theta \bar \chi_{\mu}(x)  
+ i \theta \sigma^{\nu} \bar \theta  v_{\nu \mu}(x)  
\\
\nonumber
& + & \theta \theta f_{\mu}(x) 
+ \bar \theta \bar \theta \bar h_{\mu}(x)
\\
\nonumber
 & + & \bar \theta \bar \theta \theta^\alpha \Big( \eta_{\mu\alpha}(x) - {i \over \sqrt{2}}\sigma^\nu_{\alpha \dot\alpha} \partial_\nu \bar \chi_\mu^{\dot \alpha}(x) \Big)  
\\
\nonumber
& + & \theta \theta \bar \theta_{\dot\alpha} \Big(\bar \rho_{\mu}^{\dot\alpha} (x) + {i \over \sqrt{2}}\bar \sigma^{\nu \dot \alpha \alpha} \partial_\nu \psi_{\mu\alpha}(x)\Big)  
\\
\nonumber
& + & \theta \theta \bar \theta \bar \theta \Big(c_{\mu}(x) -{1\over 4} \square a_\mu (x)\Big). 
\eeqa 
The conditions (\ref{reduced}, \ref{div}, \ref{DDV}) lead to the following relations among the component fields (x dependence suppressed):

 - bosons:
\beq
\label{i}
\partial^{\mu} a _{\mu} = 0, 
\eeq
\beq
\label{ii}
f_{\mu} = h_{\mu} = c_{\mu}  = 0,
\eeq
\beq
\label{iii}
\bar \sigma^{\mu \dot \alpha \alpha} \sigma^{\lambda}_{\alpha \dot \beta}
(\partial_{\lambda} a_{\mu} +  v_{\lambda \mu}) = 0,
\eeq
\beq
\label{iv}
\sigma^{\lambda}_{\beta \dot \alpha} \bar \sigma^{\mu \dot \alpha \alpha} 
(\partial_{\lambda} a_{\mu} -  v_{\lambda \mu}) = 0,
\eeq
\beq
\label{v}
\bar \sigma^{\mu \dot\alpha\alpha} \square a_\mu + \bar \sigma^{\mu \dot \alpha \beta}
\sigma^{\nu}_{\beta \dot \beta} \bar \sigma^{\lambda \dot \beta \alpha} \partial_{\nu} 
 v_{\lambda \mu} = 0,
\eeq
\beq
\label{vi}
\partial^{\nu}  v_{\nu \mu} = \partial^{\mu}  v_{\nu \mu} = 0.
\eeq

- fermions:

\beq
\label{I}
\partial^{\mu} \psi_{\mu \alpha} = \partial^{\mu} \bar \chi_{\mu \dot \alpha} = 0,
\eeq
\beq
\label{II}
\bar \sigma^{\mu\dot \alpha \alpha} \psi_{\mu\alpha} = 
\sigma^{\mu}_{\alpha \dot \alpha} \bar \chi^{\dot \alpha}_{\mu} = 0,
\eeq
\beq
\label{III}
\sigma^{\mu}_{\alpha \dot \alpha} \bar \sigma^{\nu \dot \alpha \beta} \partial_{\nu} \psi_{\mu\beta}
= \bar \sigma^{\mu \dot \alpha \alpha} \sigma^{\nu}_{\alpha \dot \beta}  \partial_{\nu} 
\bar \chi_{\mu}^{\dot \beta} = 0,
\eeq
\beq
\label{IV}
 \eta_{\mu}^{\alpha} = 
\bar \rho^{\dot \alpha}_{\mu} =0,
\eeq
\beq
\label{V}
\square \psi_{\mu}^{\alpha} = \square \bar \chi_{\mu}^{\dot \alpha} = 0,
\eeq

As can be seen from the above equations we can eliminate the fields $f_{\mu}$, $h_{\mu}$, $c_{\mu}$, 
$\eta_{\mu \alpha}$ and $\bar \rho_{\mu \dot \alpha}$ from the supermultiplet, so that we are left with the component fields $a_{\mu}$, $\psi_{\mu \alpha}$, $\bar \chi_{\mu \dot \alpha}$ and $ v_{\lambda \mu}$ only. In general they form several possible supermultiplets depending on their Lorenz and spin structure, they obey also corresponding equations of motion.
\vskip1cm

{\bf 3. Supermultiplets.}

 Systematic decomposition into irreducible supersymmetry representations can be performed by looking at the structure of the tensor field $ v_{\lambda \mu}$.

{\bf 3.a Symmetric case:} $\quad  v_{\lambda \mu} =  v_{\mu \lambda}$ .

   Combining the relations (\ref{iii}) and (\ref{iv}) one obtains:
\beq
\label{epsbos1}
-i \epsilon^{\nu \mu \lambda \rho}  v_{\lambda \rho} + \partial^{\nu} a^\mu - \partial^\mu a^\nu = 0.
\eeq
\beq
\label{epsbos2}
-i \epsilon^{\nu \mu \lambda \rho} \partial_{\lambda} a_\rho +  v^{\nu \mu}  -  v^{\mu\nu} = 0.
\eeq

Symmetric field $ v_{\lambda \mu}$ gives:
\beq
\partial_\nu a_\mu - \partial_\mu a_\nu = 0.
\eeq
This means that the field $a_\mu$ must be of the form:
\beq
a_\mu (x) = \partial_\mu a (x).
\eeq
Since the component fields are interrelated by the supersymmetry transformation:
\beqa
\delta_\xi a_\mu  & \equiv & (\xi Q + \bar \xi \bar Q) a_\mu = \partial_\mu (\delta_\xi a) = \sqrt{2}\xi \psi_\mu + 
\sqrt{2}  \bar \xi \bar \chi_\mu,
\\
& ... &
\eeqa
the remaining independent fields must be also of the form:
\beqa
\psi_\mu (x) = \partial_\mu \psi (x) & , & \quad \bar \chi_\mu (x) = \partial_\mu \bar \chi (x),
\\
\nonumber
 v_{\lambda \mu} (x) & = & \partial_\lambda \partial_\mu v (x),
\\
\nonumber
f_\mu (x) = \partial_\mu f (x) & , & \quad h_\mu (x) = \partial_\mu h (x),
\\
\nonumber
& ... &
\eeqa
Eqs. (\ref{i}-\ref{V}) simplify now significantly and the "symmetric" case can be summarized: the solution to the Eqs. (\ref{explicit}) consists of one chiral ($a + v,\; \psi_\alpha$) and one antichiral ($a - v, \; \bar \chi_{\dot\alpha}$) supermultiplet with two scalar fields satisfying the (massless) Klein-Gordon equations and two spinor fields satisfying the (massless)  Dirac equations. 

\beqa
\square a(x) = 0 & , &\quad \square v(x) = 0,
\\
\nonumber
\bar \sigma^{\mu\dot\alpha\alpha}\partial_{\mu} \psi_\alpha (x) = 0 & , & 
\quad
\sigma^{\mu}_{\alpha\dot\alpha}\partial_{\mu} \bar \chi^{\dot \alpha} (x) = 0 ,
\eeqa
All the remaining component fields vanish.

{\bf 3.b Antisymmetric case:} $\quad   v_{\lambda \mu} = -  v_{\mu \lambda}$ .

   In this case I can express $v^{\mu\nu}$ through  a vector field $v^\mu$:
\beq
v^{\mu\nu} = \partial^\mu v^\nu - \partial^\nu v^\mu.
\eeq
and relate it to the  tensor field 
\beq
a^{\mu\nu} = \partial^\mu a^\nu - \partial^\nu a^\mu
\eeq
 through (imaginary) duality relation:
\beq
\label{vantisym}
 v_{\nu\mu} = {i\over 4} \epsilon_{\nu\mu\lambda\rho} a^{\lambda\rho}.
\eeq
The relations (\ref{i}-\ref{vi}) can be expressed by one of these fields (eg. $a^{\mu\nu}$) and lead to the Maxwell field equations:
\beqa
\label{Maxwell}
\partial^{\nu} a_{\nu \mu} = 0, \quad \epsilon_{\nu\mu\lambda\rho} \partial^\mu a^{\lambda\rho} = 0.
\eeqa
with the Lorenz condition $\partial^{\mu} a _{\mu} = 0$. 

   Let us look now at the fermionic sector.  The component fields $\eta_{\mu}^\alpha$ and $\bar \rho_{\mu}^{\dot\alpha}$ vanish.  The vector-spin fields obey Eqs (\ref{I},\ref{II}) which means that only the spin $3/2$ is present. In addition they obey the Eqs. (\ref{III}) which can be rewritten in the form of the Rarita-Schwinger field equations:
\beqa
\label{RS}
\epsilon_{\rho\mu\lambda\nu}\bar \sigma^{\mu \dot \alpha \alpha} \partial^\lambda \psi^{\nu}_{\alpha} = 0,\quad \epsilon_{\rho\mu\lambda\nu} \sigma^{\mu}_{\alpha \dot \alpha} \partial^\lambda \bar \chi^{\nu \dot \alpha} = 0.
\eeqa
To summarize the antisymmetric case: the solution to Eqs.({\ref{explicit}) is built of two spin $3/2$ vector-spinor component fields $\psi_{\mu}^\alpha (x)$ and $\bar \chi_\mu^{\dot \alpha} (x)$ obeying the (massless) Rarita-Schwinger equations and a complex spin $1$ vector field $a_\mu (x)$ obeying the Maxwell equations. The above multiplets are sometimes called matter gravitino supermultiplets \cite{mgs}.

\vskip5mm
{\bf 4. Summary.}

Together with earlier results \cite{BH}, the "square root" of the Dirac operator $\mathcal{S}$, when acting on the superfields $F$ and $B$, gives all superfields necessary for the construction of renormalizable supersymmetric theories together with the appropriate equations of motion.

Two issues are certainly worth further study. The first concerns the masses. Even if we started from the massive Dirac equation, all resulting component fields are massless. This seems to be quite natural in supersymmetry at classical level. 

The gauge invariant interaction may be introduced in the way suggested in Ref. \cite{JS86} where the spinorial derivatives $\Dda, \Ddadot$ appearing in the operator $\mathcal S$ are 
replaced by the covariant spinorial derivatives $\mathscr{D_\alpha, \bar D_{\dot\alpha}}$:
\beq
{\mathscr{D}_{\alpha}} = \Dda + i g A_{\alpha}, \quad {\mathscr{ \bar D}_{\dot\alpha}} = \Ddadot + i g \bar A_{\dot\alpha}.
\eeq
The  above problems are currently under investigation.
\vskip1cm
{\bf 5. Acknowledgements.}

The author would like to thank Adam Bzdak, Richard Grimm and Leszek Hadasz for discussions. The hospitality of the CPT Marseille is also acknowledged.
\vskip2cm
%\bibliography{wues} \bibliographystyle{appb}

\end{document}